\documentclass[12pt,oneside]{book}
\usepackage{aobf}
\usepackage[normalem]{ulem}	

\setlist{nolistsep}

\begin{document}

\newcommand{\problem}[4]
%    {\begin{center}\textbf{#1}\end{center}
    {{\centering \textbf{#1}
    
    }    \nopagebreak

        \noindent \textit{Statement:} #2

        \noindent \textit{Source:} #3

        \noindent \textit{Remarks:} #4

        \bigskip}

\frontmatter

\title{Open Problems in \\ Analysis of Boolean Functions}
\date{Compiled for the Simons Symposium, February 5--11, 2012 \\ \bigskip For notation and definitions, see e.g.\ \texttt{http://analysisofbooleanfunctions.org}}
%\author{}
%\address{}
%\curraddr{}
%\email{}
%\thanks{}

\maketitle

\mainmatter
\setcounter{page}{2}

\problem{Correlation Bounds for Polynomials}{
    Find an explicit (i.e., in $\mathsf{NP}$) function $f \co \F_2^n \to \F_2$ such that we have the correlation bound $|\E[(-1)^{\la f(\bx), p(\bx) \ra}]| \leq 1/n$ for every $\F_2$-polynomial $p \co \F_2^n \to \F_2$ of degree at most $\log_2 n$.}{
        Folklore dating back to~\cite{Raz87,Smo87}}{
        \begin{itemize}
            \item The problem appears to be open even with correlation bound $1/\sqrt{n}$ replacing~$1/n$.
            \item Define the $\textrm{mod}_3$ function to be~$1$ if and only if the number of $1$'s in its input is congruent to~$1$ modulo~$3$.  Smolensky~\cite{Smo87} showed that $\textrm{mod}_3$ has correlation at most $2/3$ with every $\F_2$-polynomial of degree at most $c\sqrt{n}$ (where $c > 0$ is an absolute constant).  For related bounds using his techniques, there seems to be a barrier to obtaining correlation $o(1/\sqrt{n})$.
            \item Babai, Nisan, and Szegedy~\cite{BNS92} implicitly showed a function in~$\mathsf{P}$ which has correlation at most $\exp(-n^{\Theta(1)})$ with any $\F_2$-polynomial of degree at most~$.99 \log_2 n$; see also~\cite{VW08}.  Bourgain~\cite{Bou05} (see also~\cite{GRS05}) showed a similar (slightly worse) result for the $\textrm{mod}_3$ function.
        \end{itemize}}

\problem{Tomaszewski's Conjecture}{
    Let $a \in \R^n$ have $\|a\|_2 = 1$.  Then $\Pr_{\bx \sim \bn}[|\la a, \bx \ra| \leq 1] \geq 1/2$.}{
    Question attributed to Tomaszewski in~\cite{Guy89}}{
    \begin{itemize}
        \item The bound of $1/2$ would be sharp in light of $a = (1/\sqrt{2}, 1/\sqrt{2})$.
        \item Holman and Kleitman~\cite{HK92} proved the lower bound $3/8$.  In fact they proved $\Pr_{\bx \sim \bn}[|\la a, \bx \ra| < 1] \geq 3/8$ (assuming $a_i \neq \pm 1$ for all~$i$), which is sharp in light of $a = (1/2, 1/2, 1/2, 1/2)$.
    \end{itemize}}

\problem{Talagrand's ``Convolution with a Biased Coin'' Conjecture}{
    Let $f \co \bn \to \R^{\geq 0}$ have $\E[f] = 1$. Fix any $0 < \rho < 1$.  Then $\Pr[\T_\rho f \geq t] < o(1/t)$.}{
    \cite{Tal89}}{
    \begin{itemize}
        \item Talagrand in fact suggests the bound $O(\frac{1}{t\sqrt{\log t}})$.
        \item Talagrand offers a \$1000 prize for proving this.
        \item Even the ``special case'' when $f$'s domain is $\R^n$ with Gaussian measure is open.  In this Gaussian setting, Ball, Barthe, Bednorz, Oleszkiewicz, and Wolff~\cite{BBB+10} have shown the upper bound $O(\frac{1}{t\sqrt{\log t}})$ for $n = 1$ and the bound $O(\frac{\log \log t}{t\sqrt{\log t}})$ for any fixed constant dimension.
    \end{itemize}}

\problem{Sensitivity versus Block Sensitivity}{
    For any $f \btb$ it holds that $\deg(f) \leq \poly(\sens[f])$, where $\sens[f]$ is the (maximum) sensitivity, $\max_{x} |\{i \in [n] : f(x) \neq f(x^{\oplus i})\}|$.}{
        \cite{CFGS88,Sze89a,GL92,NS94}}{
        \begin{itemize}
            \item As the title suggests, it is more usual to state this as $\mathrm{bs}[f] \leq \poly(\sens[f])$, where $\mathrm{bs}[f]$ is the ``block sensitivity''.  However the version with degree is equally old, and in any case the problems are equivalent since it is known that $\mathrm{bs}[f]$ and $\deg(f)$ are polynomially related.
            \item The best known gap is quadratic (\cite{CFGS88,GL92}) and it is suggested (\cite{GL92}) that this may be the worst possible.
        \end{itemize}}

\problem{Gotsman--Linial Conjecture}{
    Among degree-$k$ polynomial threshold functions $f \btb$, the one with maximal total influence is the symmetric one $f(x) = \sgn(p(x_1 + \cdots + x_n))$, where $p$ is a degree-$k$ univariate polynomial which alternates sign on the $k+1$ values of $x_1 + \cdots + x_n$ closest to~$0$.}{
        \cite{GL94}}{
        \begin{itemize}
            \item The case $k = 1$ is easy.
            \item Slightly weaker version: degree-$k$ PTFs have total influence $O(k) \cdot \sqrt{n}$.
            \item Even weaker version: degree-$k$ PTFs have total influence $O_k(1) \cdot \sqrt{n}$.
            \item The weaker versions are open even in the case $k = 2$.  The $k = 2$ case may be related to the following old conjecture of Holzman: If $g \btR$ has degree~$2$ (for $n$~even), then $g$ has at most $\binom{n}{n/2}$ local strict minima.
            \item It is known that bounding total influence by $c(k) \cdot \sqrt{n}$ is equivalent to a bounding $\delta$-noise sensitivity by $O(c(k)) \cdot \sqrt{\delta}$.
            \item The ``Gaussian special case'' was solved by Kane~\cite{Kan09}.
            \item The best upper bounds known are $2n^{1-1/2^k}$ and $2^{O(k)} \cdot n^{1-1/O(k)}$~\cite{DHK+10}.
        \end{itemize}}

\problem{Polynomial Freiman--Ruzsa Conjecture (in the $\F_2^n$ setting)}{
    Suppose $\emptyset \neq A \subseteq \F_2^n$ satisfies $|A + A| \leq C|A|$.  Then $A$ can be covered by the union of $\poly(C)$ affine subspaces, each of cardinality at most~$|A|$.}{
        Attributed to Marton in \cite{Ruz93}; for the $\F_2^n$ version, see e.g.~\cite{Gre05}}{
        \begin{itemize}
            \item The following conjecture is known to be equivalent:  Suppose $f \co \F_2^n \to \F_2^n$ satisfies $\Pr_{\bx, \by}[f(\bx) + f(\by) = f(\bx + \by)] \geq \eps$, where $\bx$ and $\by$ are independent and uniform on~$\F_2^n$.  Then there exists a linear function $f \co \F_2^n \to \F_2^n$ such that $\Pr[f(\bx) = \ell(\bx)] \geq \poly(\eps)$.
            \item The PFR Conjecture is known to follow from the \textbf{Polynomial Bogolyubov Conjecture}~\cite{GT09a}: Let $A \subseteq \F_2^n$ have density at least~$\alpha$. Then $A+A+A$ contains an affine subspace of codimension $O(\log(1/\alpha))$.  One can slightly weaken the Polynomial Bogolyubov Conjecture by replacing $A+A+A$ with $kA$ for an integer $k > 3$.  It is known that any such weakening (for fixed finite $k$) is enough to imply the PFR Conjecture.
            \item Sanders~\cite{San10} has the best result in the direction of these conjectures, showing that if $A \subseteq \F_2^n$ has density at least $\alpha$ then $A+A$ contains $99\%$ of the points in a subspace of codimension $O(\log^4(1/\alpha))$, and hence $4A$ contains all of this subspace.  This suffices to give the Freiman--Ruzsa Conjecture with $2^{O(\log^4 C)}$ in place of $\poly(C)$.
            \item Green and Tao~\cite{GT09a} have proved the Polynomial Freiman--Ruzsa Conjecture in the case that $A$ is monotone.
         \end{itemize}}

\problem{Mansour's Conjecture}{
    Let $f \btb$ be computable by a DNF of size $s > 1$ and let $\eps \in (0,1/2]$.  Then $f$'s Fourier spectrum is $\eps$-concentrated on a collection $\calF$ with $|\calF| \leq s^{O(\log(1/\eps))}$.}{
        \cite{Man94}}{
        \begin{itemize}
            \item Weaker version: replacing $s^{O(\log(1/\eps))}$ by $s^{O_\eps(1)}$.
            \item The weak version with bound $s^{O(1/\eps)}$ is known to follow from the Fourier Entropy--Influence Conjecture.
            \item Proved for ``almost all'' polynomial-size DNF formulas (appropriately defined) by Klivans, Lee, and Wan~\cite{KLW10}.
            \item Mansour~\cite{Man95} obtained the upper-bound $(s/\eps)^{O(\log \log(s/\eps) \log(1/\eps))}$.
         \end{itemize}}

\problem{Bernoulli Conjecture}{
    Let $T$ be a finite collection of vectors in $\R^n$.  Define $b(T) = \E_{\bx \sim \bn}[\max_{t \in T} \la t, \bx \ra]$, and define $g(T)$ to be the same quantity except with~$\bx \sim \R^n$ Gaussian.  Then there exists a finite collection of vectors $T'$ such that $g(T') \leq O(b(T))$ and $\forall t \in T\ \exists t' \in T'\ \|t-t'\|_1 \leq O(b(T))$.}{
        \cite{Tal94a}}{
        \begin{itemize}
            \item The quantity $g(T)$ is well-understood in terms of the geometry of $T$, thanks to Talagrand's majorizing measures theorem.
            \item Talagrand offers a \$5000 prize for proving this, and a \$1000 prize for disproving it.
         \end{itemize}}

\problem{Fourier Entropy--Influence Conjecture}{
    There is a universal constant $C$ such that for any $f \btb$ it holds that $\bH[\wh{f}^2] \leq C \cdot \Tinf[f]$, where $\bH[\wh{f}^2] = \sum_{S} \wh{f}(S)^2 \log_2\frac{1}{\wh{f}(S)^2}$ is the spectral entropy and $\Tinf[f]$ is the total influence.}{
        \cite{FK96b}}{
        \begin{itemize}
            \item Proved for ``almost all'' polynomial-size DNF formulas (appropriately defined) by Klivans, Lee, and Wan~\cite{KLW10}.
            \item  Proved for symmetric functions and functions computable by read-once decision trees by O'Donnell, Wright, and Zhou~\cite{OWZ11}.
            \item An explicit example showing that $C \geq 60/13$ is necessary is known. (O'Donnell, unpublished.)
            \item Weaker version: the ``Min-Entropy--Influence Conjecture'', which states that there exists~$S$ such that $\wh{f}(S)^2 \geq 2^{-C \cdot \Tinf[f]}$.  This conjecture is strictly stronger than the KKL Theorem, and is implied by the KKL Theorem in the case of monotone functions.
        \end{itemize}}

\problem{Majority Is Least Stable Conjecture}{
    Let $f \btb$ be a linear threshold function, $n$~odd.  Then for all $\rho \in [0,1]$, $\Stab_\rho[f] \geq \Stab_\rho[\maj_n]$.}{
        \cite{BKS99}}{
        \begin{itemize}
            \item Slightly weaker version: If $f$ is a linear threshold function then $\NS_\delta[f] \leq \frac{2}{\pi} \sqrt{\delta} + o(\sqrt{\delta})$.
            \item The best result towards the weaker version is Peres's Theorem~\cite{Per04}, which shows that every linear threshold function~$f$ satisfies $\NS_\delta[f] \leq \sqrt{\frac{2}{\pi}} \sqrt{\delta} + O(\delta^{3/2})$.
            \item By taking $\rho \to 0$, the conjecture has the following consequence, which is also open: Let $f \btb$ be a linear threshold function with $\E[f] = 0$.  Then $\sum_{i=1}^n \wh{f}(i)^2 \geq \frac{2}{\pi}$.  The best known lower bound here is $\half$, which follows from the Khinchine--Kahane inequality; see~\cite{GL94}.
        \end{itemize}}

\problem{Optimality of Majorities for Non-Interactive Correlation Distillation}{
    Fix $r \in \N$, $n$ odd, and $0 < \eps < 1/2$.  For $f \btb$, define $P(f) = \Pr[f(\by^{(1)}) = f(\by^{(2)}) = \cdots f(\by^{(r)})]$, where $\bx \sim \bn$ is chosen uniformly and then each $\by^{(i)}$ is (independently) an $\eps$-noisy copy of~$\bx$.  Is it true that $P(f)$ is maximized among odd functions~$f$ by the Majority function $\maj_k$ on \emph{some} odd number of inputs~$k$? }{
        \cite{MO05} (originally from 2002)}{
        \begin{itemize}
            \item It is possible (e.g., for $r = 10$, $n = 5$, $\eps = .26$) for neither the Dictator ($\Maj_1$) nor full Majority ($\Maj_n$) to be maximizing.
        \end{itemize}}

\problem{Noise Sensitivity of Intersections of Halfspaces}{
    Let $f \btb$ be the intersection (AND) of $k$ linear threshold functions.  Then $\NS_\delta[f] \leq O(\sqrt{\log k}) \cdot \sqrt{\delta}$.}{
        \cite{KOS02}}{
        \begin{itemize}
            \item The bound  $O(k) \cdot \sqrt{\delta}$ follows easily from Peres's Theorem and is the best known.
            \item The ``Gaussian special case'' follows easily from the work of Nazarov~\cite{Naz03}.
            \item An upper bound of the form $\polylog(k) \cdot \delta^{\Omega(1)}$ holds if the halfspaces are sufficiently ``regular''~\cite{HKM10}.
        \end{itemize}}

\problem{Non-Interactive Correlation Distillation with Erasures}{
    Let $f \btb$ be an unbiased function.  Let $\bz \sim \{-1,0,1\}^n$ be a ``random restriction'' in which each coordinate $\bz_i$ is (independently) $\pm 1$ with probability $p/2$ each, and~$0$ with probability $1-p$.  Assuming $p < 1/2$ and $n$ odd, is it true that $\E_{\bz}[|f(\bz)|]$ is maximized when $f$~is the majority function? (Here we identify $f$ with its multilinear expansion.)}{
        \cite{Yan04}}{
        \begin{itemize}
            \item For $p \geq 1/2$, Yang conjectured that $\E_{\bz}[|f(\bz)|]$ is maximized when $f$ is a dictator function; this was proved by O'Donnell and Wright~\cite{OW12}.
            \item Mossel~\cite{MOS10} shows that if $f$'s influences are assumed at most~$\tau$ then $\E_{\bz}[|f(\bz)|] \leq \E_{\bz}[|\Maj_n(\bz)|] + o_\tau(1)$.
        \end{itemize}}

\problem{Triangle Removal in $\F_2^n$}{
    Let $A \subseteq \F_2^n$.  Suppose that $\eps 2^n$ elements must be removed from~$A$ in order to make it ``triangle-free'' (meaning there does not exist $x, y, x+y \in A$).  Is it true that $\Pr_{\bx, \by}[\bx, \by, \bx+\by \in A] \geq \poly(\eps)$, where $\bx$ and~$\by$ are independent and uniform on~$\F_2^n$?}{
        \cite{Gre05a}}{
        \begin{itemize}
            \item Green~\cite{Gre05a} showed the lower bound $1/(2\!\uparrow\!\uparrow\!\eps^{-\Theta(1)})$.
            \item Bhattacharyya and Xie~\cite{BX10} constructed an~$A$ for which the probability is at most roughly $\eps^{3.409}$.
        \end{itemize}}

\problem{Subspaces in Sumsets}{
    Fix a constant $\alpha > 0$.  Let $A \subseteq \F_2^n$ have density at least~$\alpha$.  Is it true that $A+A$ contains a subspace of codimension $O(\sqrt{n})$?}{
        \cite{Gre05a}}{
        \begin{itemize}
            \item The analogous problem for the group $Z_N$ dates back to Bourgain~\cite{Bou90}.
            \item By considering the Hamming ball $A = \{x : |x| \leq n/2 - \Theta(\sqrt{n})\}$, it is easy to show that codimension $O(\sqrt{n})$ cannot be improved.  This example is essentially due to Ruzsa~\cite{Ruz93}, see~\cite{Gre05a}.
            \item The best bounds are due to Sanders~\cite{San10a}, who shows that $A+A$ must contain a subspace of codimension $\lceil n/(1+ \log_2(\frac{1-\alpha}{1-2\alpha}))\rceil$.  Thinking of $\alpha$ as small, this means a subspace of \emph{dimension} roughly $\frac{\alpha}{\ln 2} \cdot n$.  Thinking of $\alpha = 1/2 - \eps$ for $\eps$ small, this is codimension roughly $n/\log_2(1/\eps)$.  In the same work Sanders also shows that if $\alpha \geq 1/2 - .001/\sqrt{n}$ then $A+A$ contains a subspace of codimension~$1$.
            \item As noted in the remarks on the Polynomial Freiman--Ruzsa/Bogolyubov Conjectures, it is also interesting to consider the relaxed problem where we only require that $A+A$ contains $99\%$ of the points in a large subspace.  Here it might be conjectured that the subspace can have codimension $O(\log(1/\alpha))$.
        \end{itemize}}

\problem{Aaronson--Ambainis Conjecture}{
    Let $f \btI$ have degree at most $k$.  Then there exists $i \in [n]$ with $\Inf_i[f] \geq (\Var[f]/k)^{O(1)}$.}{
        \cite{Aar08,AA11}}{
        \begin{itemize}
            \item True for $f \btb$; this follows from a result of O'Donnell, Schramm, Saks, and Servedio~\cite{OSSS05}.
            \item The weaker lower bound $(\Var[f]/2^k)^{O(1)}$ follows from a result of Dinur, Kindler, Friedgut, and O'Donnell~\cite{DFKO07}.
        \end{itemize}}

\problem{Bhattacharyya--Grigorescu--Shapira Conjecture}{
    Let $M \in \F_2^{m \times k}$ and $\sigma \in \{0,1\}^k$.  Say that $f \co \F_2^n \to \zo$ is \emph{$(M,\sigma)$-free} if there does not exist $X = (x^{(1)}, \dots, x^{(k)})$ (where each $x^{(j)} \in \F_2^n$ is a row vector) such that $MX = 0$ and $f(x^{(j)}) = \sigma_j$ for all $j \in [k]$. Now fix a (possibly infinite) collection $\{(M^1, \sigma^1), (M^2, \sigma^2), \cdots \}$ and consider the property $\calP_n$ of functions $f \co \F_2^n \to \{0,1\}$ that $f$ is $(M^i,\sigma^i)$-free for all $i$.  Then there is a one-sided error, constant-query property-testing algorithm for~$\calP_n$.}{
        \cite{BGS10}}{
        \begin{itemize}
            \item The conjecture is motivated by a work of Kaufman and Sudan~\cite{KS08} which proposes as an open research problem the characterization of testability for linear-invariant properties of functions $f \co \F_2^n \to \{0,1\}$.  The properties defined in the conjecture are linear-invariant.
            \item Every property family $(\calP_n)$ defined by $\{(M^1, \sigma^1), (M^2, \sigma^2), \cdots \}$-freeness is \emph{subspace-hereditary}; i.e., closed under restriction to subspaces.  The converse also ``essentially'' holds.~\cite{BGS10}.
            \item For $M$ of rank one, Green~\cite{Gre05a} showed that $(M,1^k)$-freeness is testable. He conjectured this result extends to arbitrary~$M$; this was confirmed by Kr\'{a}l', Serra, and Vena~\cite{KSV08} and also Shapira~\cite{Sha09}. Austin~\cite{Sha09} subsequently conjectured that  $(M,\sigma)$-freeness is testable for arbitrary~$\sigma$; even this subcase is still open.
            \item The conjecture is known to hold when all $M^i$ have rank one~\cite{BGS10}. Also, Bhattacharyya, Fischer, and Lovett~\cite{BFL12} have proved the conjecture in the setting of~$\F_p$ for affine constraints $\{(M^1,\sigma^1), (M^2,\sigma^2), \dots\}$ of ``Cauchy--Schwarz complexity'' less than~$p$.
        \end{itemize}}

\problem{Symmetric Gaussian Problem}{
    Fix $0 \leq \rho, \mu, \nu \leq 1$.  Suppose $A,B \subseteq \R^n$ have Gaussian measure $\mu$, $\nu$ respectively.  Further, suppose $A$ is centrally symmetric: $A = -A$.  What is the minimal possible value of $\Pr[\bx \in A, \by \in B]$, when $(\bx, \by)$ are $\rho$-correlated $n$-dimensional Gaussians?}{
        \cite{CR10}}{
        \begin{itemize}
            \item It is equivalent to require both $A = -A$ and $B = -B$.
            \item Without the symmetry requirement, the minimum occurs when $A$ and $B$ are opposing halfspaces; this follows from the work of Borell~\cite{Bor85}.
            \item A reasonable conjecture is that the minimum occurs when $A$ is a centered ball and $B$ is the complement of a centered ball.
        \end{itemize}}

\problem{Standard Simplex Conjecture}{
    Fix $0 \leq \rho \leq 1$.  Then among all partitions of $\R^n$ into $3 \leq q \leq n+1$ parts of equal Gaussian measure, the maximal noise stability at $\rho$ occurs for a ``standard simplex partition''.  By this it is meant a partition $A_1, \dots, A_q$ satisfying $A_i \supseteq \{x \in \R^n : \la a_i, x \ra  > \la a_j, x \ra\ \forall j \neq i\}$, where $a_1, \dots, a_q \in \R^n$ are unit vectors satisfying $\la a_i, a_j \ra = -\frac{1}{q-1}$ for all $i \neq j$. Further, for $-1 \leq \rho \leq 0$ the standard simplex partition minimizes noise stability at~$\rho$.}{
        \cite{IM09}}{
        \begin{itemize}
            \item Implies the Plurality Is Stablest Conjecture of Khot, Kindler, Mossel, and O'Donnell~\cite{KKMO04}; in turn, the Plurality Is Stablest Conjecture implies it for $\rho \geq -\frac{1}{q-1}$.
        \end{itemize}}

\problem{Linear Coefficients versus Total Degree}{
    Let $f \btb$.  Then $\sum_{i=1}^n \wh{f}(i) \leq \sqrt{\deg(f)}$.}{
        Parikshit Gopalan and Rocco Servedio, ca.\ 2009}{
        \begin{itemize}
            \item More ambitiously, one could propose the upper bound $k \cdot \binom{k-1}{\frac{k-1}{2}} 2^{1-k}$, where $k = \deg(f)$.  This is achieved by the Majority function on~$k$ bits.
            \item Apparently, no bound better than the trivial $\sum_{i=1}^n \wh{f}(i) \leq \Tinf[f] \leq \deg(f)$ is known.
        \end{itemize}}

\problem{$k$-wise Independence for PTFs}{
    Fix $d \in \N$ and $\eps \in (0,1)$.  Determine the least $k = k(d,\eps)$ such that the following holds: If $p \co \R^n \to \R$ is any degree-$d$ multivariate polynomial, and $\bX$ is any $\R^n$-valued random variable with the property that each $\bX_i$ has the standard Gaussian distribution and each collection $\bX_{i_1}, \dots, \bX_{i_k}$ is independent, then $|\Pr[p(\bX) \geq 0] - \Pr[p(\bZ) \geq 0]| \leq \eps$, where $\bZ$ has the standard $n$-dimensional Gaussian distribution.}{
        \cite{DGJ+09}}{
        \begin{itemize}
            \item For $d = 1$, Diakonikolas, Gopalan, Jaiswal, Servedio, and Viola~\cite{DGJ+09} showed that $k = O(1/\eps^2)$ suffices.  For $d = 2$, Diakonikolas, Kane, and Nelson~\cite{DKN10} showed that $k = O(1/\eps^8)$ suffices.  For general $d$, Kane~\cite{Kan11b} showed that $O_d(1) \cdot \eps^{-2^{O(d)}}$ suffices and that $\Omega(d^2/\eps^2)$ is necessary.
        \end{itemize}}

\problem{$\eps$-biased Sets for DNFs}{
    Is it true for each constant $\delta > 0$ that $s^{-O(1)}$-biased densities $\delta$-fool size-$s$ DNFs?  I.e., that if $f \co \{0,1\}^n \to \{-1,1\}$ is computable by a size-$s$ DNF and $\vphi$ is an $s^{-O(1)}$-biased density on $\{0,1\}$, then $|\E_{\bx \sim \{0,1\}^n}[f(\bx)] - \E_{\by \sim \vphi}[f(\by)]| \leq \delta$.}{
        \cite{DETT10}, though the problem of pseudorandom generators for bounded-depth circuits dates back to~\cite{AW85}}{
        \begin{itemize}
            \item De, Etesami, Trevisan, and Tulsiani~\cite{DETT10} show the result for $\exp(-O(\log^2(s) \log \log s))$-biased densities.  If one assumes Mansour's Conjecture, their result improves to $\exp(-O(\log^2 s))$.  More precisely, they show that $\exp(-O(\log^2(s/\delta) \log \log (s/\delta)))$-biased densities $\delta$-fool size-$s$ DNF.  They also give an example showing that $s^{-O(\log(1/\delta))}$-biased densities are \emph{necessary}.  Finally, they show that $s^{-O(\log(1/\delta))}$-biased densities suffice for read-once DNFs.
        \end{itemize}}

\problem{PTF Sparsity for Inner Product Mod 2}{
    Is it true that any PTF representation of the inner product mod~$2$ function on $2n$ bits, $\IP_{2n} \co \F_2^{2n} \to \{-1,1\}$, requires at least~$3^n$ monomials?}{
        Srikanth Srinivasan, 2010}{
        \begin{itemize}
            \item Rocco Servedio independently asked if the following much stronger statement is true:  Suppose $f, g \btb$ require PTFs of sparsity at least $s, t$, respectively; then $f \oplus g \co \bits^{2n} \to \bits$ (the function $(x,y) \mapsto f(x)g(y)$) requires PTFs of sparsity at least $st$.
        \end{itemize}}

\problem{\sout{Servedio--Tan--Verbin Conjecture}}{
    Fix any $\eps > 0$. Then every monotone $f \btb$ is $\eps$-close to a $\poly(\deg(f))$-junta.}{
        Elad Verbin (2010) and independently Rocco Servedio and Li-Yang Tan (2010)}{
        \begin{itemize}
            \item One can equivalently replace degree by decision-tree depth or maximum sensitivity.
            \item RESOLVED (in the negative) by Daniel Kane, 2012.
        \end{itemize}}

\problem{Average versus Max Sensitivity for Monotone Functions}{
    Let $f \btb$ be monotone.  Then $\Tinf[f] < o(\sens[f])$.}{
        Rocco Servedio, Li-Yang Tan, 2010}{
        \begin{itemize}
            \item The tightest example known has $\Tinf[f] \approx \sens[f]^{.61}$; this appears in a work of O'Donnell and Servedio~\cite{OS08b}.
        \end{itemize}}

\problem{Approximate Degree for Approximate Majority}{
    What is the least possible degree of a function $f \co \bits^n \to [-1, -2/3] \cup [2/3, 1]$ which has $f(x) \in [2/3,1]$ whenever $\sum_{i=1}^n x_i \geq n/2$ and has $f(x) \in [-1,-2/3]$ whenever $\sum_{i=1}^n x_i \leq -n/2$?}{
        Srikanth Srinivasan, 2010}{
        \begin{itemize}
            \item Note that $f(x)$ is still required to be in $[-1, -2/3] \cup [2/3, 1]$ when $-n/2 < \sum_{i=1}^n x_i < n/2$.
        \end{itemize}}

\problem{Uncertainty Principle for Quadratic Fourier Analysis}{
    Suppose $q_1, \dots, q_m \co \F_2^n \to \F_2$ are polynomials of degree at most~$2$ and suppose the indicator function of $(1, \dots, 1) \in \F_2^n$, namely $\AND \co \F_2^n \to \{-1,1\}$, is expressible as $\AND(x) = \sum_{i=1}^m c_i (-1)^{q_i(x)}$ for some real numbers~$c_i$. What is a lower bound for~$m$?}{
        Hamed Hatami, 2011}{
        \begin{itemize}
            \item Hatami can show that $m \geq n$ is necessary but conjectures $m \geq 2^{\Omega(n)}$ is necessary.  Note that if the $q_i$'s are of degree at most~$1$ then $m = 2^n$ is necessary and sufficient.
            \item The \emph{Constant-Degree Hypothesis} is a similar conjecture made by Barrington, Straubing, and Th\'{e}rien~\cite{BST90} in 1990 in the context of finite fields.
        \end{itemize}}

%2 percolation problems
%singularity problem for random matrices
%submodularity
%conjectures from Hamed talk
%beating $2^{quadratic}$, even 2-eps for the Friedgut version of the min-entropy influence
%http://www-personal.umich.edu/~mishlie/DoubleBubbles.pdf
%gianchi, elchanan-neeman for the robust isoperimetry

\backmatter
\newpage
\bibliographystyle{alpha}
\bibliography{../bib/odonnell-bib}

\end{document}